\documentclass[conference]{IEEEtran}
\IEEEoverridecommandlockouts
\usepackage{multirow}
\usepackage{subfig}
\usepackage{svg}

\usepackage{cite}

\ifCLASSINFOpdf
  \usepackage[font=footnotesize]{caption}
\else
\fi
\usepackage{algorithm}
\usepackage{algpseudocode}

\usepackage{array}
\usepackage{amsmath}

\usepackage{fixltx2e}
\usepackage{tikz}

\newcommand\copyrighttext{%
  \footnotesize \copyright  2023 IEEE.  Personal use of this material is permitted.  Permission from IEEE must be obtained for all other uses, in any current or future media, including reprinting/republishing this material for advertising or promotional purposes, creating new collective works, for resale or redistribution to servers or lists, or reuse of any copyrighted component of this work in other works.}
\newcommand\mycopyrightnotice{%
\begin{tikzpicture}[remember picture,overlay]
\node[anchor=south,yshift=10pt] at (current page.south) {\fbox{\parbox{\dimexpr\textwidth-\fboxsep-\fboxrule\relax}{\copyrighttext}}};
\end{tikzpicture}%
}

\begin{document}

%
\title{Active Learning Based Domain Adaptation for Tissue Segmentation of Histopathological Images}



%
\author{\IEEEauthorblockN{Saul Fuster\IEEEauthorrefmark{3}\textsuperscript{\IEEEauthorrefmark{1}},
Farbod Khoraminia\IEEEauthorrefmark{2}\textsuperscript{\IEEEauthorrefmark{1}},
Trygve Eftestøl\IEEEauthorrefmark{3},
Tahlita C.M. Zuiverloon\IEEEauthorrefmark{2},
Kjersti Engan\IEEEauthorrefmark{3}}
\smallskip
\IEEEauthorblockA{\IEEEauthorrefmark{3}Dept. of Electrical Engineering and Computer Science, University of Stavanger, 4021 Stavanger, Norway}
\IEEEauthorblockA{\IEEEauthorrefmark{2}Dept. of Urology, Erasmus MC Cancer Institute, University Medical Center, 3015 GD Rotterdam, The Netherlands}
\thanks{*Joint first authorship. SF: methodology, software, experiments, drafting, writing. FK: data collection, labelling, evaluation, editing.}}


\maketitle
\mycopyrightnotice

\begin{abstract}
Accurate segmentation of tissue in histopathological images can be very beneficial  for defining regions of interest (ROI) for streamline of diagnostic and prognostic tasks. Still, adapting to different domains is essential for histopathology image analysis, as the visual characteristics of tissues can vary significantly across datasets. Yet, acquiring sufficient annotated data in the medical domain is cumbersome and time-consuming. The labeling effort can be significantly reduced by leveraging active learning, which enables the selective annotation of the most informative samples. Our proposed method allows for fine-tuning a pre-trained deep neural network using a small set of labeled data from the target domain, while also actively selecting the most informative samples to label next. We demonstrate that our approach performs with significantly fewer labeled samples compared to traditional supervised learning approaches for similar F1-scores, using barely a 59\% of the training set. We also investigate the distribution of class balance to establish annotation guidelines.
\end{abstract}


%
\IEEEpeerreviewmaketitle

\section{Introduction}

\begin{figure*}[t]
    \centering
    \includegraphics[trim=4cm 1.5cm 4cm 1.5cm, width=\textwidth, height=9.8cm]{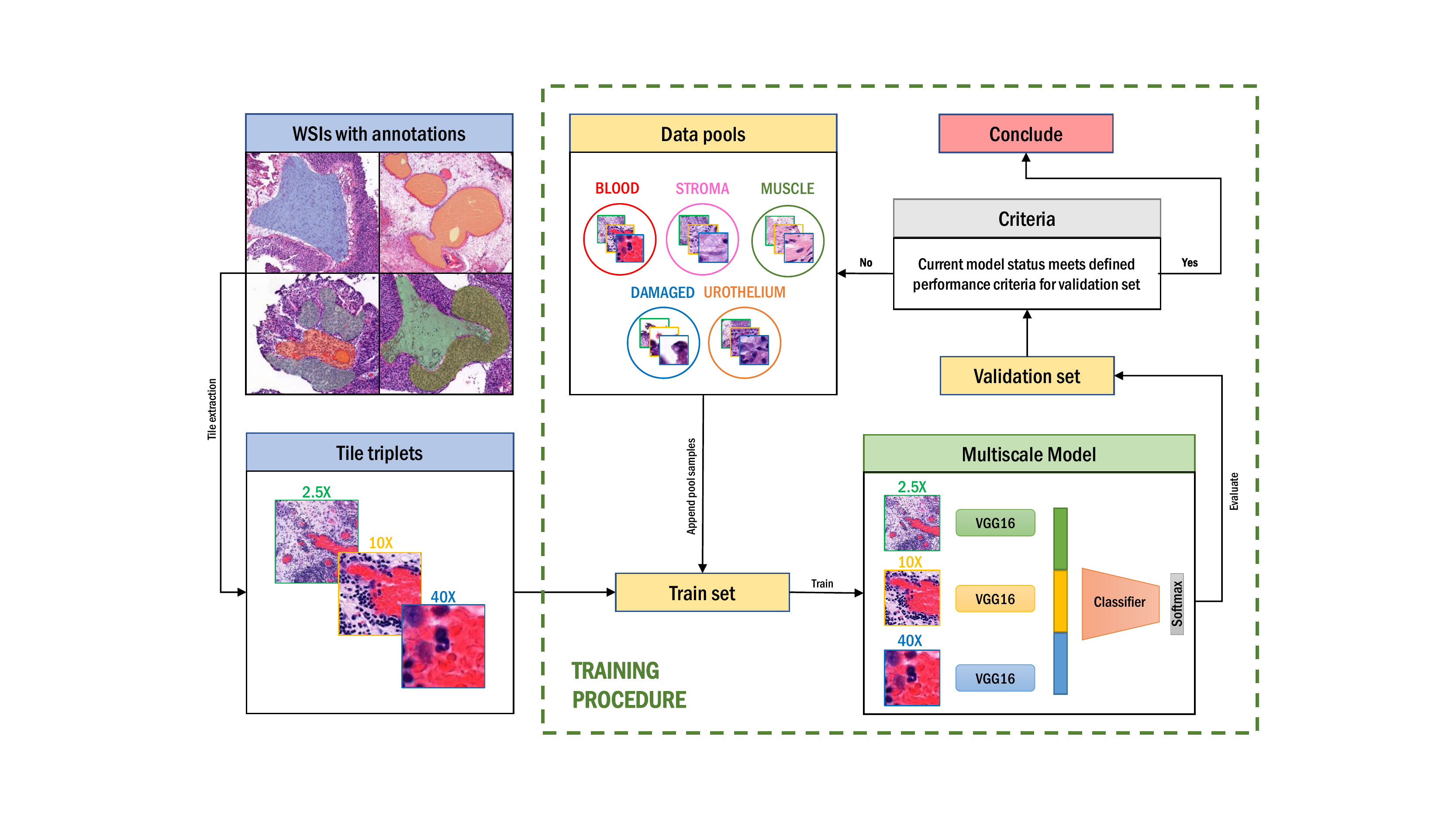}
    \caption{Active learning framework TRI-AL. Pathologists annotate ROIs, from which tile triples are extracted at different magnifications levels (2.5x, 10x, 40x) to form a triplet. During the training stage, an starting training set is defined for training a multiscale model. The model performance is evaluated on the validation set. Then, stopping criteria decides to resume with another training iteration or conclude the learning. In case that the criteria is not met, new samples from the pools of data are drawn and appended to the current version of the train set.}
    \label{fig:train_inference}
\end{figure*}

Computer-aided diagnosis (CAD) systems that utilize machine learning techniques for medical imaging analysis have been shown to be an effective way to reduce subjectivity and speed up the diagnostic process \cite{van2021deep}. Digital microscopy scanners are capable of generating high-resolution digital images from scanned tissue sections, also known as Whole Slide Images (WSI). These images can are pre-stored at various magnification levels, allowing pathologists to adjust the zoom level like they would with physical microscopes. Lower magnification is typically used to view tissue-level morphology, while higher magnification is useful for examining cell-level features. 

Bladder cancer is among the most commonly diagnosed cancer types. According to the World Health Organization, over 573,000 new cases and 213,000 deaths were registered in 2020 \cite{Sung2021GlobalCS}. WSIs from bladder cancer are highly disorganized scanned tissue sections for several reasons. Urothelial carcinomas often present papillary structures, elongated finger-like bundles of tissue, that alter the normal appearance of the urothelial lining. Also, transurethral resection of a bladder tumour (TURBT) is a complicated operation that difficults a clean tumour extraction as the cauterization process leaves damaged tissue areas. As a result, a significant amount of artifacts and other non-diagnostically-relevant tissue is present within the slides. Using all the regions of a WSI for model training or inference as input would often add unnecessary noise. Manual annotations of potential regions of interests (ROIs) is an arduous, time-consuming and labor-intensive task. Hence automatic methods have emerged to reduce this time constraint. Tissue segmentation in computational pathology (CPATH) enables the analysis of specific ROIs within a WSI and can also improve the performance of the model by actively selecting the most informative tissue types \cite{budd2021survey, komura2018machine}. A Multiscale Approach for Whole-Slide Image Segmentation of five Tissue Classes in Urothelial Carcinoma Slides \cite{wetteland2020multiscale} proposes a multiscale convolutional neural network (CNN) that can effectively segment five different tissue classes in non-muscle invasive bladder urothelial carcinoma slides. The model classifies all input areas into blood, damaged, muscle, stroma and urothelium tissue. They demonstrate that their approach outperforms existing methods and is able to handle the large size and variability of WSIs within their private cohort. The presence of a domain shift between images obtained from different laboratories impedes the performance of deep learning models on out-of-distribution samples \cite{stacke2019closer}. Thus, implementing the algorithm on a new dataset may require of an adaptation. However, the cost of labeling resources is critical, especially in the clinic.

When dealing with limited data, it is frequently necessary to resort to the implementation of deep learning methodologies that are more cost-efficient, such as active learning (AL). AL is a variant of supervised learning (SL) involving human interaction during training, also referred to as having a human in the loop during training \cite{settles2009active}. The goal of an AL setup is to actively interfere during the training procedure, extract new data points from the classes the model struggles to comprehend and append them to the training set \cite{ren2021survey}. Therefore, an human in the loop is expected to spend time resources for labeling data points from the class the model demands. Recent works within the field of CPATH have demonstrated implementing AL techniques results in similar performances to SL, with constrained data settings \cite{yang2017suggestive, doyle2011active, smailagic2018medal, jin2021reducing, li2021pathal, gal2017deep, lutnick2019integrated}. Analogously, AL can enhance the tissue segmentation model from \cite{wetteland2020multiscale} by selecting informative samples from an undetermined dataset with limited annotations. Despite the publication of protocols for critical bias assessment of clinical models, no official guidelines have been established regarding the required numbers of annotations, images, and laboratories to capture the variation present in real-world data \cite{collins2021protocol, morales2021artificial}. The need for more well-defined class sampling strategies within the field of histopathology arises. Due to privacy concerns, medical data cohorts often cannot be made publicly available. This leads to limited model predictive generalization as medical applications are developed for a target dataset. Consequently, domain adaptation of deep learning models into unexplored data domains is needed.

In this work, we propose a domain adaptation framework for deep learning models within the field of histopathology. We choose to adapt an algorithm that segments bladder cancer WSIs into different tissue types. The model architecture uses a multiscale CNN backbone that incorporates information from different magnification levels, which has been developed using another dataset from a different hospital. In order to adapt the model to a new unseen domain, we adopt an AL strategy for a more efficient labeling effort. We proactively select samples to be included into the training data based on preemptive results on the validation set. We show that the proposed AL approach is more profitable and can be integrated to reduce labeling costs. On top of that, we also aim to guide pathologists in which annotations to provide deep learning models before investing substantial amounts of effort. We estimate a balance between class distribution and model performance, validated using a small initial subset of annotations. By assessing the model's initial performance, pathologists can then proceed with further annotations guided by this intuition.

\section{Material and Methods}
\subsection{Dataset}
We have collected a set of high-risk non-muscle invasive bladder cancer (HR-NMIBC) WSI from the first TURBT from a multi-centre cohort provided by Erasmus MC, Rotterdam, The Netherlands. WSIs were stained with Haematoxylin and Eosin (H\&E) and scanned using a 3DHistech P1000 scanner at 80x magnification stored as MRXS files. The total number of slides is 155, for which a pathologist has annotated the slides with tissue types. Data heterogeneity produces more generalizing models than using higher amounts of data from the same slides \cite{jin2021reducing}.  Thus, a pathologist was asked to annotate areas in a rough, imprecise manner in order to obtain several annotated regions per WSI, within a time limit. The time usage was limited to a maximum of one hour per WSI, including diagnostic labels not used in this work. As a result, some regions were annotated in every WSI, but not the entirety of present tissue was annotated. The aim of this annotation protocol was to collect diverse scenarios, hence capture the tissue heterogeneity characteristic from bladder cancer. Moreover, to avoid incorporating a human in the loop during training, we preemptively collected available tissue type annotations and defined pools of data to draw from. In total, 127, 16, and 12 WSIs were annotated and used for training, validation and test, respectively, where the split is done on WSI level to avoid cross-contamination. Tiles from annotated regions of urothelium, stroma, muscle, blood and damaged tissue were extracted from the EMC cohort, using the strategy proposed in \cite{wetteland2020multiscale} where more details can be found. In short; tiles were extracted at 2.5x, 10x and 40x magnifications, thus forming a triplet, and a maximum of 500 triplets were extracted per WSI. A tile size of 128$\times$128 was used, and all three tiles in a triplet share a common physical point as the center pixel. Therefore, tiles at a lower magnification cover a larger physical area. The total number of tiles across sets is stated in Table \ref{tab:tileseg}.

\begin{table}[h]
\centering
\begin{tabular}{c|c|c|c|}
\cline{2-4}
                                          & \textbf{Train} & \textbf{Val} & \textbf{Test} \\ \hline
\multicolumn{1}{|l|}{\textbf{WSI}}        & 127            & 16                  & 12          \\  \hline
\multicolumn{1}{|l|}{\textbf{Blood}}      & 35105          & 2500                & 4106          \\ 
\multicolumn{1}{|l|}{\textbf{Damaged}}    & 65920          & 2500                & 57296         \\ 
\multicolumn{1}{|l|}{\textbf{Muscle}}     & 67007          & 2500                & 50347         \\ 
\multicolumn{1}{|l|}{\textbf{Stroma}}     & 86978          & 2500                & 79338         \\ 
\multicolumn{1}{|l|}{\textbf{Urothelium}} & 91006          & 2500                & 38813         \\ \hline
\multicolumn{1}{|l|}{\textbf{Total}}      & 346016         & 12500               & 229900        \\ \hline
\end{tabular}
\caption{Number of WSI and tiles per tissue type (class) in train/val/test split of the available dataset of high-risk non-muscle invasive bladder cancer.}
\label{tab:tileseg}
\end{table}

\subsection{Model Architecture}
The tissue segmentation model TRI-CNN proposed in \cite{wetteland2020multiscale} is adopted into our pipeline. This model was trained using WSIs of NMIBC patients from Stavanger University Hospital (SUS), Stavanger, Norway. The model architecture consists of a multiscale CNN setup that aggregates local and global information. Triplets are fed through three weight-independent VGG16 backbones trained for each of the magnifications, concatenating each output feature vectors before classification. Then, the formed feature vector is fed through the classifier to predict a tissue class, using softmax activation. A representation of the model architecture is presented in Fig. \ref{fig:train_inference}.

\subsection{Active Learning Procedure}
Active learning (AL) can be a powerful tool for improving the performance of deep learning models when labeled data is limited \cite{ren2021survey}. In AL, the model is initially trained on a small labeled dataset. Thereafter, based on intermediate performance metrics and a query strategy for requesting additional samples of the most informative class, a human annotator is asked to label a small number of additional samples. The model then uses this newly labeled data to update its parameters and improve its performance. This process is repeated until either the model reaches a satisfactory level of performance, the resources are exhausted or a set number of iterations has been conducted. One of the main advantages of AL is that it can significantly reduce the amount of labeled data required to train a deep learning model to achieve acceptable performance. Additionally, active learning can also improve the model's performance by allowing it to adapt to changing conditions or concepts over time.
However, AL also has some limitations. It requires human intervention, which can be time-consuming or cumbersome to organize, and it may introduce bias into the training process. Overall, the choice between SL and AL will depend on the specific needs of the application and the availability of labeled data. 

AL strategies, or query strategies, involve analyzing the information value of unlabeled instances. The most commonly used query strategy is uncertainty sampling \cite{settles2009active, yang2016active}. This framework revolves around the model uncertainty in labeling certain instances, revolving in probabilistic methods, such as entropy and prediction confidence. Predictions on unlabeled instances presenting a high degree of entropy are selected for inclusion to the training data. These are later transferred to a domain expert for labeling. Inspired by these methodologies, but with the goal of domain adaptation, our sampling strategy falls under the uncertainty query strategies. For the AL procedure, we further divided the training set, $\mathcal{X}$, into initial training subset $\mathcal{X}_0$ and pools of training data $\mathcal{P}$. The algorithm selects new samples based on model performance. Considering the false negative ratio (FNR) on the validation set $\mathcal{V}$ and total sample size $S$, $N^i_j$ samples from the pools $\mathcal{P}$ are appended into the current training set, for every class $i$ on iteration $j$. Furthermore, we define a relative class balance size $\delta^j_i$ to represent the percentage of data points for class $i$ on the current training set $\mathcal{X}_j$. These steps are iteratively repeated until $\mathcal{P}$ cannot provide more samples of the requested class or $J$ iterations have passed, see Algorithm \ref{alg:altrain} for more details.

\begin{algorithm}[H]
  \caption{Active Learning Train Procedure}
  \begin{algorithmic}[1]
    \Procedure{ALtrain}{$\mathcal{X}, \mathcal{V}, J, S$}
    \State $\mathcal{X}_0, \mathcal{P} \gets \mathcal{X}$\Comment{Split train set into initial subset and pools}
       \While{$j < J$ \textbf{or} $len(\mathcal{P}_i) > N^i_j$}
        \For{$j \gets 1$ to $J$}
          \State $MODEL_{TRAIN}(\mathcal{X}_j)$
          \State $MODEL_{EVAL}(\mathcal{V})$
          \For{$i \gets 1$ to $I$}\Comment{Per class new data points}
            \State $N^{i}_{j} = \frac{FNR^{i}_{j-1}\cdot S}{\sum_{i}^{}FNR^{i}_{j-1}}$
            \State $\mathcal{X}^i_j \overset{+}{\leftarrow} sample(\mathcal{P}, N^{i}_{j})$
          \EndFor
        \EndFor
       \EndWhile
    \EndProcedure
  \end{algorithmic}
\label{alg:altrain}
\end{algorithm}


\begin{table*}[]
\centering
\begin{tabular}{l|wc{5em}||wc{5em}|wc{5em}|wc{5em}|wc{5em}|wc{5em}||wc{5em}|wc{5em}|wc{5em}|}
\cline{2-10}
                       & TRI-CNN \cite{wetteland2020multiscale} & TRI-SL(20\%) & TRI-SL(40\%) & TRI-SL(60\%) & TRI-SL(80\%) & TRI-SL(100\%) & \textbf{TRI-AL\textsubscript{ITER}} & \textbf{TRI-AL\textsubscript{OOD}} & \textbf{TRI-AL\textsubscript{ENT}}\\ \cline{2-10} \hline
\multicolumn{1}{|l|}{\textbf{Blood}}         & 29.06(-)      & 53.05(2.33)      & \textbf{58.46(2.50)}      & 55.21(3.24)      & 53.31(3.06)      & 56.22(2.34)      & 55.36(3.50)      & 54.44(2.15)      & 51.84(2.94)\\ 
\multicolumn{1}{|l|}{\textbf{Damaged}}       & 75.65(-)      & 89.28(0.39)      & 91.82(0.23)      & \textbf{92.77(0.27)}      & 91.55(0.66)      & 92.44(0.31)      & 92.65(0.22)      & 92.12(0.17)      & 92.63(0.31)\\ 
\multicolumn{1}{|l|}{\textbf{Muscle}}        & 71.54(-)      & 89.67(0.49)      & 89.37(0.56)      & 92.11(0.71)      & 92.51(0.64)      & 92.04(0.70)      & \textbf{92.69(1.15)}      & 92.37(0.35)      & 92.11(0.57)\\ 
\multicolumn{1}{|l|}{\textbf{Stroma}}        & 70.73(-)      & 87.59(0.29)      & 86.93(0.76)      & 88.31(0.92)      & 88.03(1.06)      & \textbf{88.47(0.64)}      & 88.40(1.24)      & 88.07(0.49)      & 87.41(0.58)\\ 
\multicolumn{1}{|l|}{\textbf{Urothelium}}    & 82.45(-)      & 90.48(0.75)      & 91.89(0.18)      & 92.66(0.30)      & 92.20(0.64)      & 93.41(0.28)      & \textbf{93.45(0.46)}      & 93.19(0.09)      & 93.39(0.27)\\ \hline
\multicolumn{1}{|l|}{\textbf{Total (micro)}} & 73.06(-)      & 88.01(0.30)      & 88.92(0.54)      & 90.08(0.41)      & 89.64(0.53)      & 90.23(0.30)      & \textbf{90.34(0.66)}      & 89.95(0.23)      & 89.69(0.41)\\ 
\multicolumn{1}{|l|}{\textbf{Total (macro)}} & 65.89(-)      & 82.01(0.63)      & 83.70(0.71)      & 84.21(0.68)      & 83.52(0.78)      & \textbf{84.52(0.46)}      & 84.51(0.97)      & 84.04(0.42)      & 83.48(0.74)\\ \hline
\end{tabular}
\caption{F1 scores per tissue type. We compare the model pretrained TRI-CNN on another dataset, to supervised learning (TRI-SL) approaches with varying training data sizes, to entropy-based active learning (TRI-AL\textsubscript{ENT}) and our proposed active learning models (TRI-AL\textsubscript{ITER}, TRI-AL\textsubscript{OOD}).}
\label{tab:f1table}
\end{table*}

\section{Experiments}
All experiments are done using transfer learning from TRI-CNN, using the same hyperparameters as in the original algorithm \cite{wetteland2020multiscale}. All models were trained using unfrozen VGG16s, a multi-class cross entropy loss function, learning rate of 1.5e-4 and SGD optimizer. Early stopping was enabled for interrupting the training if there was no improvement in the validation loss for five consecutive epochs. The models were trained five times in order to determine the mean and standard deviation of the results.

We evaluate the performance of TRI-CNN on our test set; then, we compare it to the proposed SL and AL strategies using data from Erasmus MC, Rotterdam, The Netherlands. For our experiments, we compare the performance of standard supervised learning (TRI-SL) on different sampling of the training set, in order to leverage the performance at various thresholds of dataset sizes. Based on the entire available training set, we trained models using a fraction of the data, namely from 20 to 100\%, in intervals of 20\%. As for the AL methods, we define a stopping criteria based on resource exhaustion and a fixed number of iterations. The initial training set $\mathcal{X}_0$ is made of 25000 triplets per class. Every iteration $i$, 20000 new samples $S$ are added to training data $\mathcal{X}_j$. We defined a maximum of 5 iterations for training the model TRI-AL\textsubscript{ITER}, while the model TRI-AL\textsubscript{OOD} refers to the model trained until a class pool runs out of data points. For comparison matters, we trained an algorithm TRI-AL\textsubscript{ENT} based on entropy uncertainty. In this case, 30000 triplets from the pools $\mathcal{P}$ are sampled based on class distribution $\delta$ of the entire training set $\mathcal{X}$. From these, the top 20000 triplets with the highest entropy are selected.

\section{Results \& Discussion}
Results in terms of F1 scores are presented in Table \ref{tab:f1table}. First, we observe that TRI-CNN, although achieves a respectable performance, falls short in comparison to the models trained on the new cohort, thus showing that domain adaptation is needed. TRI-CNN achieved a micro F1 score of 73.06. Regarding TRI-SL models, the performance improved along with the size of the dataset, considering that the best SL model is that using the entire training set. Nevertheless, AL strategies achieve higher performance using data more efficiently. The query strategy in TRI-AL\textsubscript{ITER} reaches a micro average F1 score of 90.34, compared to 90.23 for SL using 100\% of the training samples. The AL data accounts for 59\% of the available training samples. It is also worth mentioning the disparity in terms of performance for the blood tissue class in regards to other tissue types. One of the reasons refers to regular missclassifications. Most of the blood annotations are those of vessels present within the stroma, so its only natural that blood can be misinterpreted as stroma. However, the main reason is that the number of blood samples in the test set for blood is heavily limited in comparison to the other classes. Naturally, a minor number of blood false positives results in a significant downgrade in terms of metrics. Focusing on the TRI-AL\textsubscript{ITER} model performance results, blood has a recall of 86.31 while for stroma is 86.77, barely a 0.66 difference.

\begin{figure}[]
\centering
\includegraphics[width=\columnwidth, height=6.2cm]{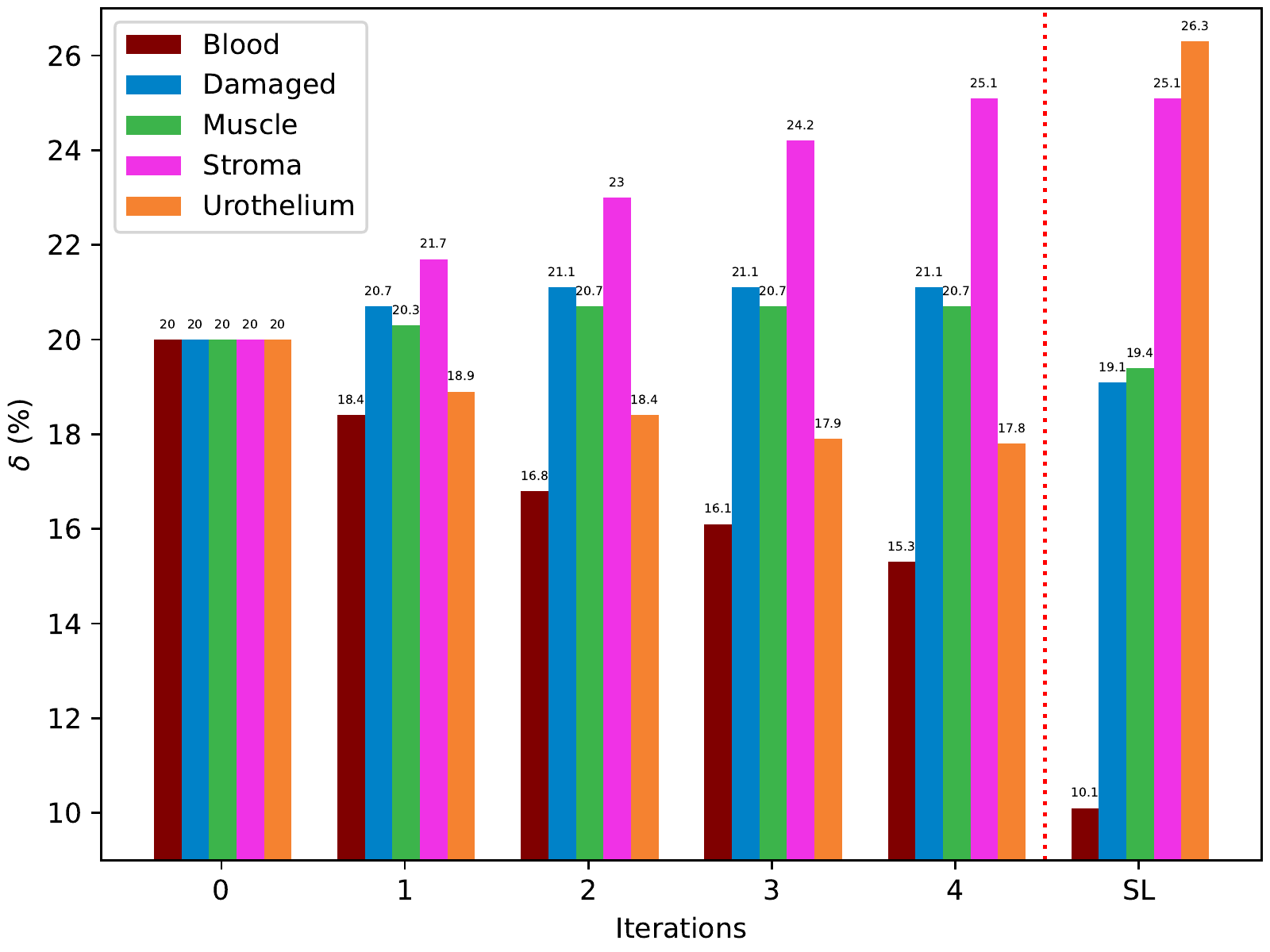} 
\caption{Class relative size per class $\delta_i$ over the iterations for TRI-AL\textsubscript{ITER} model. We observe that the algorithm prioritizes harder classes over simpler ones as the iterations pass. As $\mathcal{X}_j$ increases, $\delta_i$ reaches an asymptote for damaged, muscle and urothelium tissue classes, while penalizing blood in favor of stroma. Class distribution does not match that annotated from a pathologist, as per the TRI-SL section.}
\label{fig:barplot}
\end{figure}

The development of the training set class distribution for TRI-AL\textsubscript{ITER} in comparison to the ones used for TRI-SL can be seen as described in Fig \ref{fig:barplot}. Looking at the relative data balance per class, we identify a tendency over the iterations where damaged, muscle and urothelium class distribution $\delta$ reach an asymptote. Also, we observe that the algorithm deems to suggest that stroma is significantly harder to learn and requires a more extensive number of data points in comparison to other classes. Even then, the margin for $\delta$ progressively slows down and we would expect it to become stagnant.

Regarding the comparison of the iteration TRI-AL\textsubscript{ITER} and data exhaustion TRI-AL\textsubscript{OOD} query strategies, we noticed that allowing the algorithm to add samples until a class pool $\mathcal{P}_i$ is exhausted does not translate into overall better performance, although that might be different in case that the pool was larger. To further evaluate the performance of our model TRI-AL\textsubscript{ITER}, we compare it to entropy-based class instance selection strategies TRI-AL\textsubscript{ENT}. Our results indicate that, on average, TRI-AL\textsubscript{ITER} outperforms the entropy-based approach TRI-AL\textsubscript{ENT}, both in terms of per-class and total aggregated metrics, indicating that it is better at distinguishing between different tissue types. However, it should be noted that we also observed lower variation in the performance of the entropy-based models between runs, as indicated by the standard deviation metrics. This suggests that the entropy-based approach may be more consistent and less susceptible to fluctuations in performance across different training iterations. Nevertheless, it is important to note that the entropy-based query strategy is more computationally expensive, as it requires the calculation of sampled data points. In contrast, our proposed method does not require any additional computation beyond the training of the model.

\begin{figure*}[t]
    \centering
    \includegraphics[trim=0.3cm 5cm 0.3cm 5cm, clip, width=2\columnwidth]{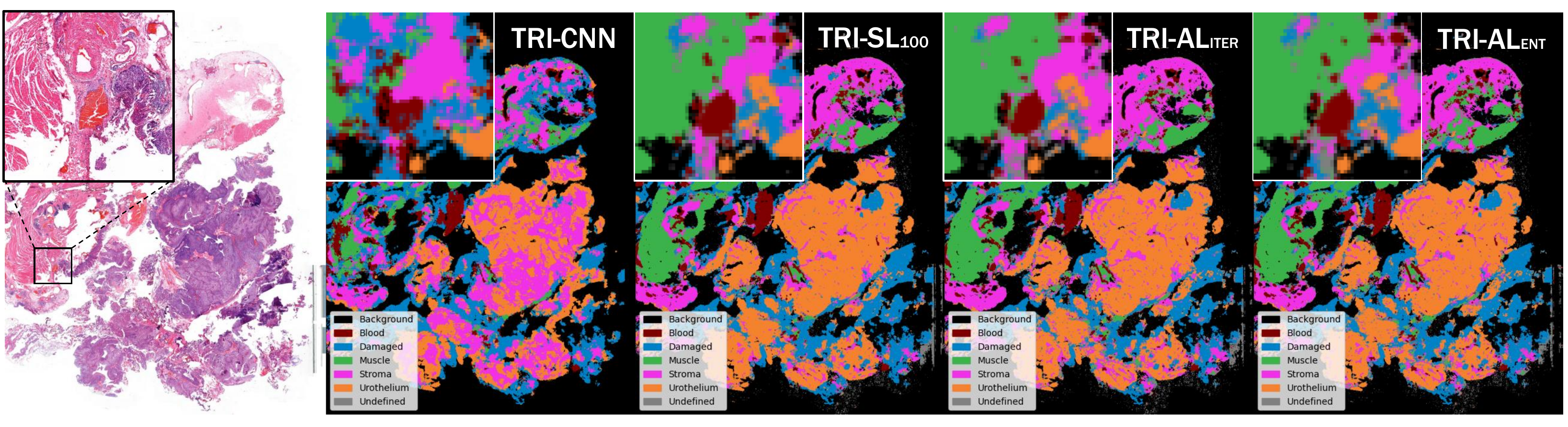}
    \caption{Segmentation prediction of a WSI using the active learning model at inference stage. Areas from all tissue types are extracted for analysis at the inference stage, resulting in a labelled colormap based on tile predictions. The models trained in the new cohort can discern different tissue types accurately.}
    \label{fig:test_example}
\end{figure*}

Pathologists also visually inspected the model's segmentation results by overlaying the predicted masks over the raw WSIs. Illustrative examples of the model's segmentation results for a representative WSI of the test set are shown in Fig. \ref{fig:test_example} for a ROI and the entire WSI, respectively. Upon visual inspection of the results by experts, according to TRI-CNN segmentation, we have identified four main observations: staining effects leading to false positives of blood in regions with high levels of eosin stain, non-cauterized damaged areas such as blur or folding, the risk of misinterpreting infiltrative immune cells as urothelial cells, and the potential for the model to predict urothelium with significant cytoplasm as stroma. As per the models trained in the new cohort (TRI-SL, TRI-AL), it was confirmed that these models accurately segmented and classified different tissue types.

\section{Conclusion \& Future Work}
In this work, we proposed a active learning  framework with a multiscale CNN for domain adaptation of a tissue segmentation model of bladder cancer histopathological images. Our proposed method achieved a F1 score of 90.34 using 59\% of the training data, and outperformed supervised learning strategies that used all available samples. Our results suggest that active learning can be an effective strategy for reducing the labeling effort in histopathological image analysis. Regarding domain adaptation, we observed that we were able to customize the model to the new domain using a small labeled set. Moreover, we also presented a suggested data annotation per class burden for tissue segmentation of bladder cancer WSI. Furthermore, this model can be introduced as a pre-processing step for other applications that require tissue segmentation for ROI extraction.


\section*{Acknowledgment}
This research has received funding from the European Union's Horizon 2020 research and innovation program under grant agreements 860627 (CLARIFY).



\bibliographystyle{IEEEtran}
\bibliography{refs}
%
%
%

\end{document}